\documentclass[letterpaper]{PoS}

\PoS{PoS(HEP2005)400}
\usepackage{epsfig}
\usepackage{amssymb}
\usepackage{times}

\newcommand{\urll}[1]{\href{#1}{#1}}%
\newcommand{\had}{parallel session talk}                                    

\newcommand{\etal}{{\em et al.}}

\newcommand{\gevcc}{\hbox{ GeV}\!/\!c^2}
\newcommand{\gev}{\hbox{ GeV}}

\newcommand{\mev}{\hbox{ MeV}}

\newcommand{\kev}{\hbox{ keV}}

\def\ltap{\mathop{\raisebox{-.4ex}{\rlap{$\sim$}} 
\raisebox{.4ex}{$<$}}}

\newcommand{\cfrac}[2]{\textstyle \frac{#1}{#2}}

\newcommand{\slj}[3]{\mbox{$^{#1}${\ifcase#2\or S\or 
	 P\or D\or F\or G\fi}$_{#3}$}}
\newcommand{\sLj}[3]{{}^{#1}\!#2_{#3}}
\newcommand{\jpsi}{\ensuremath{J\!/\!\psi}}

\def\slashii#1{\setbox0=\hbox{$#1$}             
   \dimen0=\wd0                                 
   \setbox1=\hbox{\sl/} \dimen1=\wd1            
   \ifdim\dimen0>\dimen1                        
      \rlap{\hbox to \dimen0{\hfil\sl/\hfil}}   
      #1                                        
   \else                                        
      \rlap{\hbox to \dimen1{\hfil$#1$\hfil}}   
      \hbox{\sl/}                               
   \fi}                                         %
\title{Hadronic Physics and Exotics}

\ShortTitle{Hadronic Physics \& Exotics}
\author{\speaker{Chris Quigg}\thanks{Fermilab is operated by Universities Research Association Inc.\ under
Contract No.\ DE-AC02-76CH03000 with the U.S.\ Department of Energy.}\\
        Theoretical Physics Department \\ Fermi National Accelerator 
        Laboratory\\
P.O. Box 500, Batavia, Illinois USA 60510-0500\\
        E-mail: \email{quigg@fnal.gov}}

\abstract{I report on the state of hadronic physics and spectroscopy 
as reflected in contributions to the HEP2005 Europhysics Conference. 
Topics of interest include lattice field theory calculations of the 
hadron spectrum, the continuing quest to account for the proton's spin, 
pentaquarks, high-statistics Dalitz-plot analyses, excited 
charmed--strange mesons, quarkonium spectroscopy, and the new levels 
associated with the charmonium spectrum.
I also reflect on the goals of hadronic physics and assess the 
current state of the art.

\begin{flushright}
    \textsf{FERMILAB--CONF--05--356--T}
\end{flushright}}

\FullConference{International Europhysics Conference on High Energy Physics\\
		 July 21st - 27th 2005\\
		 Lisboa, Portugal}

\begin{document}

\section{Hadronic Physics Self-Assessment}
Over the past three years, the discovery of many new states  and 
 remarkably incisive explorations of a broad range of phenomena have 
 renewed interest in hadronic physics and spurred many lively 
 conversations between theory and experiment. It seems appropriate, 
 when the subject is in a healthy state of ferment, to begin with a 
 brief assessment of the value and aspirations of hadronic physics.

Hadron phenomenology and spectroscopy does not test the standard model.
We have a qualitative understanding of QCD phenomenology, but many
aspects are not calculable from first principles.  While we may learn
how to refine our approximations to QCD, much analysis of experimental
information relies on highly stylized, truncated pictures of the
implications of the theory.  We make models for new (and old!)  states:
approximations such as potential models, or intuitive pictures of
substructure.  The competing pictures are not mutually exclusive;
quantum superpositions are possible.  We will never discard QCD as the
theory of the strong interactions if these pictures fail for the next
state we find.

 These are fair observations, and they merit our serious attention.
 I would note that there is value to both fundamental and applied
 science, and that the apparently less glamourous work of applied
 science may be just what we need to get at the fundamental lessons. 
 Moreover, exploration---the task of discovering what phenomena exist 
 and of developing systematics---helps us to understand what the 
 fundamental questions are, and how we might best address them. It 
 was, after all, the tension among the quark model of hadrons, the 
 parton-model description of deeply inelastic scattering, and the 
 nonobservation of free quarks that led us to quantum chromodynamics. 
 The construction of a crossing-symmetric, Regge-behaved amplitude 
 for linearly rising trajectories was a foundational event in string 
 theory~\cite{Veneziano:1968yb}.

 Physics doesn't advance by perturbation theory alone, and it is worth
 recalling that one of QCD's signal achievements is explaining what
 sets the mass of the proton---or, if you like, what accounts for
 nearly all the visible mass of the Universe.  The insight that the
 mass of the proton arises from the energy stored up in confining three
 quarks in a small volume, not from the masses of the constituents
 themselves, is a landmark in our understanding of
 Nature~\cite{Wilczek:1999be}.  The value of that insight isn't
 diminished because it is a little bit qualitative, or because a
 quantitative execution of the idea requires the heavy machinery of
 lattice field theory\footnote{See Ref.~\cite{alltalks} for
 contributions to the Hadronic Physics parallel
 sessions.}~\cite{heller,ukawa}.

 More generally, there is great value in a convincing physical picture
 that can show us the way to an answer (whether or not precise and
 controlled), or show that some tempting simplifying assumptions are
 unwarranted.  The chiral quark model~\cite{Manohar:1983md}, which
 identifies the significant degrees of freedom on the 1-GeV scale as
 constituent quarks and Goldstone bosons, offers a nice example.  It
 points to the $u$-$d$ asymmetry in the light-quark sea of the
 proton~\cite{Eichten:1991mt}, and predicts a negative polarization of
 the strange (but not antistrange) sea, casting doubt on a seemingly
 harmless assumption that underlies the Ellis--Jaffe sum
 rule~\cite{Ellis:1973kp}.  A lifetime of staring at
 $\mathcal{L}_{\mathrm{QCD}}$ wouldn't lead to these expectations.

 We can value \textit{anschaulich} explanations as sources of intuition
 and instruments of exploration, while keeping clearly in mind their
 limitations, as we try to address many open-ended questions, including:
 What is a hadron?  What are the apt degrees of freedom?  What
 symmetries are fruitful?  What are the implications of QCD under
 extreme conditions?
 
  Emergent behavior---in the form of phenomena that are not simply
  derived from the underlying microphysics---is, moreover, quite
  ubiquitous in particle physics, and especially in hadronic physics.
  For example, as QCD becomes strongly coupled at low energies, new
  phenomena emerge that are not immediately obvious from the
  Lagrangian.  Confinement and chiral symmetry breaking, with the
  implied appearance of Goldstone bosons, are specific illustrations.
  A graceful description entails new degrees of freedom that may be
  expressed in a model or---in the best of cases---in a new effective
  field theory.

The synthesis of principles through dialogue with experiment is central
to the way hadronic physics is constructed, and runs through the agenda
of the parallel sessions.  I am firmly convinced that decoding hadronic
phenomena in today's experiments develops habits of mind that we will
cherish when the LHC brings surprises.

\section{Where Does the Proton's Spin Reside?}
Contributions to the parallel session reminded us that we do not have a
complete answer to the question, ``What is a proton?''  The spin of a
polarized proton may be partitioned among the quarks (and antiquarks),
gluons, and orbital angular momentum according to the expression
$\cfrac{1}{2} = \cfrac{1}{2}\Delta\Sigma + \Delta G + L_{q} + L_{g}$.
New measurements from the COMPASS experiment improve the determination
of the quark-antiquark component to $\Delta \Sigma =
0.237^{+0.024}_{-0.029}$, and anchor the gluon contribution at $x=0.1$
as $\frac{\Delta G}{G}(x_{g}=0.1) = +0.024 \pm 0.089 \pm
0.057$~\cite{pretz,procureur}.  At the same time, studies of transverse
spin effects in the HERMES experiment give evidence for orbital angular
momentum carried by the quarks~\cite{avetisyan}.  BELLE also
contributes to this program by measuring the fragmentation function of
a transversely polarized quark~\cite{seuster}.  This area offers but
one example of diverse experiments making common cause.

\section{Searching for Connections}
The essence of doing science consists in \textit{making connections} 
that lead us beyond independent explanations for distinct phenoma 
toward a coherent understanding of many phenomena. A network of 
understanding helps us see how different observations fit together 
and---very important---helps us know enough to recognize that 
something \textit{doesn't fit.}

Connections among experiments or observations are not the only
important ones.  Whenever it is possible, we need to make connections
between experimental systematics, phenomenological models, and the QCD
Lagrangian---either directly, or through effective field theories,
lattice field theory, or a controlled approximation to full QCD. I
would also stress the potential value of reaching toward connections
with our knowledge of nuclear forces and with the phenomena that occur
in nuclear matter under unusal conditions.

We recognize different circumstances under which various approximations
to QCD can be regarded as controlled expansions in small parameters.
Nonrelativistic QCD applies to heavy-heavy ($Q_{1}\bar{Q}_{2}$) mesons,
for which the quark masses greatly exceed the QCD scale parameter,
$m_{Q_{i}} \gg \Lambda_{\mathrm{QCD}}$.  Befitting its aptness for the
nonrelativistic limit, NRQCD takes as its expansion parameter $v/c$,
the heavy-quark velocity divided by the speed of light.  Heavy-quark
effective theory (HQET) applies usefully to heavy-light ($Q\bar{q}$)
systems, for which $m_{Q} \gg \Lambda_{\mathrm{QCD}}$.  In first
approximation, the spin of the heavy quark is regarded as static, so
the ``light-quark spin'' $\vec{j}_{q} = \vec{L}+ \vec{s}_{q}$ is a good
quantum number.  The relevant expansion parameter is
{$\Lambda_{\mathrm{QCD}}/m_{Q}$}.  Chiral symmetry is a valuable
starting point for light quark systems ($q_{1}\bar{q}_{2}$) with
$m_{q_{i}} \ll \Lambda_{\mathrm{QCD}}$.  In this case, the expansion
parameter compares the current-quark mass to the scale of
chiral-symmetry breaking, and is generally taken as $m_{q}/4\pi
f_{\pi}$, where $f_{\pi}$ is the pion decay constant. In a growing 
array of settings, lattice QCD embodies a controlled approximation 
that expresses the full dynamical content of the theory~\cite{davies}.

\section{Seeking the Relevant Degrees of Freedom}
Much of our insight into how hadrons behave follows from the
simplifying assumption that mesons are quark--antiquark states, baryons
are three-quark states, and that the quarks have only essential
correlations.  In the case of baryons, this reasoning leads us to the
plausible starting point of SU(6) (flavor-spin) wave functions, which
indeed offer a useful framework for discussing magnetic moments and
other static properties.  Some well-known observations, however, show us the
limitations of the zeroth-order guess.  If we examine deeply inelastic
scattering in the limit as $x \to 1$, {spin asymmetries} indicate that
the SU(6) wave functions are inadequate~\cite{Hughes:1983kf}, and the
ratio $F_{2}^{n}/F_{2}^{p}$ is far from the uncorrelated expectation of
$\cfrac{2}{3}$~\cite{Melnitchouk:1995fc}.

Under what circumstances might it be fruitful---or even essential---to
consider diquarks as physical objects~\cite{Anselmino:1992vg}?  The
algebra of SU(3)$_{c}$ tells us that the $\mathbf{3} \otimes
\mathbf{3}$ quark--quark combination is attractive in the
$\mathbf{3^{*}}$ representation that corresponds to an antisymmetric
diquark structure.  A simple analysis suggests that the attraction of
$[qq]_{\mathbf{3^{*}}}$ is half as strong as that of the
$[q\bar{q}]_{\mathbf{1}}$ ($\mathbf{3} \otimes \mathbf{3^{*}} \to
\mathbf{1}$) channel.  For many years, it has seemed to make sense to
regard members of the scalar nonet \{$f_{0}(600) = \sigma, \kappa(900),
f_{0}(980), a_{0}(980)$\} as $qq\bar{q}\bar{q}$ states organized as
$[[qq]_{\mathbf{3^{*}}}[\bar{q}\bar{q}]_{\mathbf{3}}]_{\mathbf{1}}$~\cite{Jaffe:1976ig}.
Recently, \textit{intrinsic diquarks} ($|uuudc \bar c\rangle$) and
intrinsic double-charm Fock states ($|uud c \bar c c \bar c\rangle$)
have been advanced as an explanation of the production of the SELEX
$\Xi(ccd)$ and $\Xi(ccu)$ states~\cite{Brodsky:2004hh}.  Diquarks as
objects have elicited new attention under the stimulus of experimental
evidence for pentaquark states~\cite{Jaffe:2003sg,Shuryak:2005pk,Jenkins:2004tm}.  (The
attention to pentaquarks should be seen as part of a broader investigation into
the existence of configurations, or body plans, beyond $qqq$ and $q\bar{q}$.) That work,
in turn, has led Wilczek and collaborators to revisit the
Chew--Frautschi systematics of $N,\Delta$
resonances~\cite{selem}, and to assert that it is useful to
view even low-spin, light baryons as $q[qq]_{\mathbf{3^{*}}}$
configurations.  What can lattice QCD tell us about the shape of $qqq$
baryons---both at the lowest spins and at high angular 
momenta~\cite{Cristoforetti:2004kj}?  Can
the quark--diquark picture be reconciled with intuition from the
$1/N_{c}$ expansion~\cite{'tHooft:1973jz,Dashen:1993jt}?

It is worth testing and extending the $q[qq]_{\mathbf{3^{*}}}$ proposal
by considering its implications for doubly heavy ($QQq$) baryons.  The
comparison with heavy-light ($Q\bar{q}$) mesons offers a chance to
calibrate the attractive forces in the $\mathbf{3^{*}}$ and
color-singlet channels~\cite{ebert,Ebert:2005xj}.  Similarly, extending
studies of the systematics of $qq \cdot \bar{q}\bar{q}$ states to $Qq
\cdot \bar{Q}\bar{q}$ states should, over the long term, develop and
challenge the way we think about diquarks.  Finally, in heavy-ion
collisions, we should be alert for tests of the utility of diquarks in
color--flavor locking, color superconductivity, and other novel
phenomena.  Tugging the diquark concept this way and that will help
elucidate the value of colorspin~\cite{Jaffe:1976ih} as an organizing
principle for hadron spectroscopy, and help us understand the relevance
of color-nonsinglet spectroscopy~\cite{Jaffe:2005md}.  Similar in their
aspirations are the considerations of diquark--triquark
configurations~\cite{Karliner:2003dt} and of the power of the
chiral-soliton picture for baryon
spectroscopy~\cite{Diakonov:1997sj,Diakonov:1997mm,Ellis:2004uz}.

\section{Exotic Baryons (Pentaquarks)}
Over the past three years, numerous experiments have reported 
evidence for narrow exotic baryons carrying quantum numbers incompatible 
with the standard $qqq$ body plan.\footnote{See Volker Burkert's summary at 
Uppsala~\cite{burkert} for a recent survey of the evidence.} These 
reports include many sightings of $\Theta^{+}(\approx1540)$, with 
$K^{+}n$ quantum numbers; a recent claim of $\Theta^{++}(1530)$, with 
$K^{+}p$ quantum numbers, in the STAR experiment at 
RHIC~\cite{kabana}; evidence for $\Xi^{--,0}(1862)$ and their 
antiparticles in the NA49 experiment at CERN; and evidence for a 
baryon with negative charm, $\Theta_{c}^{0}(3099)$, in the H1 
experiment at DESY. All of these states could be interpreted as 
$qqqq\bar{q}$ pentaquarks, and those composed of light quarks $u,d,s$ 
alone could be assigned to a $\mathbf{10^{*}}$ representation of flavor 
SU(3). It is by no means obvious on dynamical grounds that narrow 
pentaquarks should populate full multiplets. Quantitative information 
about pentaquarks---or other states that lie beyond the 1960s quark 
model,\footnote{Exotic $qq\bar{q}\bar{q}$ ``tetraquarks,'' for 
example~\cite{klausf}.} but within the spectrum allowed by quantum 
chromodynamics---would allow us to refine heuristic pictures of hadron 
structure and sharpen our understanding of QCD in the confinement limit. 
Accordingly, the pentaquark candidates have elicited much 
theoretical attention.

Many sensitive, high-resolution experiments do not support the 
observation of pentaquarks. Indeed, \textit{no claim is unchallenged,}
and it is hard to argue that every experiment---whether offering 
positive or negative evidence---is both significant and
correctly interpreted.

Recently, the two experiments that began the pentaquark rush have 
reported new data. The LEPS experiment~\cite{nakanobeijing} has taken 
new runs on liquid hydrogen and liquid deuterium targets. In the 
reaction $\gamma d \to K^{-}pX$, they see excesses in the mass 
spectrum of particles recoiling against $K^{-}p$ at $1.53\gev$ and 
$1.6\gev$; at the lower peak, the ratio of signal to
$\sqrt{\mathrm{signal}+\mathrm{background}}$ is approximately 5. The 
CLAS experiment at JLab has taken data on hydrogen and deuterium with 
samples about an order of larger than in their original experiment. 
In $\gamma p \to K_{S}K^{+}n$ with approximately 1500 counts per
4-MeV bin, they see no $\Theta^{+}$ signal~\cite{devita}. There is 
also no sign of $\Theta^{+}$ in their $\gamma d \to K^{-}pK^{+}n$ 
sample~\cite{burkert}; an increased estimate of background reduces 
the significance of their original claim to  $\approx 3\sigma$.

Here in Lisbon, we have heard limits on pentaquark production in
$Z^{0}$ decays from the DELPHI experiment~\cite{pukhaeva}; the
nonobservation of $\Theta^{+}, \Xi^{--}$ in the HERA-B
experiment~\cite{zivko}; and status reports on the contending evidence
on strange and charmed pentaquarks from the $e^{\pm}p$ collider
experiments H1 and ZEUS~\cite{chekanov,ozerov}. The $B$-factory 
experiments BaBar and Belle reported limits on pentaquark production 
based on the study of interactions with detector 
elements~\cite{grauges,mizuk}, a lovely  technique.

The case for exotic baryons remains unproved. If you wonder how it 
could be possible for an apparently robust signal, confirmed in multiple 
experiments, to prove misleading, I refer you to the great sensation 
of the 1969 Lund Conference, the two-peak structure of the split-$A_{2}$ 
(now $a_{2}$) meson~\cite{lund69}. 

\section{Dalitz-Plot Analyses}
Among many parallel-session contributions on production and decay
dynamics, I would like to point to three applications of Dalitz-plot
techniques that are representative of a new era in the extraction of
decay amplitudes and relative phases.  CLEO-$c$ reports a large number
of studies in progress~\cite{selen}.  Among them, the aim of
determining the strong phase between the decays $D^{0} \to
K^{\pm}K^{*\mp}$ for extraction of $\phi_{3}=\gamma$ from $B^{\pm} \to
K^{\pm}K^{*\mp}K^{\pm}$ has a direct practical application.
	
In BaBar's study of the $D^{0} \to \bar{K}^{0}K^{+}K^{-}$ Dalitz plot,
the dominant channels are seen to be $D^{0} \to \bar{K}^{0}a_{0}(980),
\bar{K}^{0}\varphi, K^{-}a_{0}^{+}(980)$~\cite{altenburg}.  The
amplitude information offers new possibilities for studying the scalar
nonet, which is also a target of the KLOE studies of $e^{+}e^{-} \to
\varphi \to \gamma f_{0}(980), \gamma a_{0}(980)$~\cite{gauzzi}.  KLOE
has also examined 5-$\gamma$ and 7-$\gamma$ final states in
the reaction $e^{+}e^{-} \to \varphi \to \gamma\eta$.  
Their noteworthy results include a measurement of the slope parameter in the 
$\pi^{0}\pi^{0}\pi^{0}$ channel and a determination of the branching fraction
$\mathcal{B}(\eta \to \pi^{0}\gamma\gamma) = (8.4 \pm 2.7 \pm 1.4)
\times 10^{-5}$, about ten times smaller than a 1984 GAMS result, and 
in line with chiral perturbation theory
	
\section{Beyond Idealizations}
There is potentially great value to be gained by stretching our models
and calculations beyond the domains in which we first encountered them.
By leaving the comfort zone, we may happen on effects that were
unimportant---or concealed---in the original setting.  An excellent
example is the prospect of extending our descriptions of the
$\psi\;(c\bar{c})$ and $\Upsilon\;(b\bar{b})$ systems to the spectrum
of $B_{c}\;(b\bar{c})$ mesons~\cite{Eichten:1994gt}.  
Several factors contribute to the theoretical interest in $B_{c}$.  The
$b\bar{c}$ system interpolates between heavy-heavy ($Q\bar{Q}$) and
heavy-light ($Q\bar{q}$) systems.  The unequal-mass kinematics and the
fact that the charmed quark is more relativistic in a $b\bar{c}$ bound
state than in the corresponding $c\bar{c}$ level imply an enhanced 
sensitivity to effects beyond nonrelativistic quantum mechanics.

The new element in $b\bar{c}$ theory is lattice QCD calculations that
include dynamical quarks.  A Glasgow--Fermilab collaboration predicts
$M(B_{c}) = 6304 \pm 20\mev$~\cite{Allison:2004be}.  Establishing the
$B_{c}$ ground state in nonleptonic decays---$\pi\jpsi, a_{1}\jpsi$ are
the most promising final states---will pin down the mass with greater
certainty than is possible in the semileptonic $\jpsi\ell\nu$ channel.
A first measurement by the CDF experiment in the $\jpsi\,\pi$ channel
gives $M(B_{c})=6287 \pm 5\mev$~\cite{Acosta:2005us}, in pleasing
agreement with the lattice computation.  Beginning to reconstruct some
part of the $b\bar{c}$ spectrum in $\gamma$ or $\pi^{+}\pi^{-}$
cascades to the ground state will be an experimental
\textit{tour-de-force.}

Let us take a moment to review some elementary points about 
meson taxonomy that are relevant to intermediate cases---including 
the $b\bar{c}$ system. Two useful classification schemes are familiar in 
atomic spectroscopy as the $LS$ and $jj$ coupling schemes. Any state 
can be described in any scheme, through appropriate configuration 
mixing, but it is prudent to keep in mind that a choice of basis can 
guide---or maybe misguide---our thinking.

For equal-mass meson systems ($q\bar{q}$ or $Q\bar{Q}$) it is
traditional to couple the orbital angular momentum, $\vec{L}$, with the
total spin of the quark and antiquark, $\vec{S} =
\vec{s}_{q}+\vec{s}_{\bar{q}}$.  This is the standard practice for
light mesons, and is now familiar for the designation of quarkonium
($c\bar{c}$ and $b\bar{b}$) levels.  The good quantum numbers are
then $S$, $L$, and $J$, with $\vec{J} = \vec{L} + \vec{S}$, and we
denote the spin-singlet and spin-triplet levels as \slj{1}{1}{0} --
\slj{3}{1}{1}; \slj{1}{2}{1} -- \slj{3}{2}{0,1,2}; \slj{1}{3}{2} --
\slj{3}{3}{1,2,3}; and, in general, as $\sLj{1}{L}{L}$ --
$\sLj{3}{L}{L-1,L,L+1}$.

In the case of heavy-light ($Q\bar{q}$) mesons, it is suggestive to
couple the difficult-to-flip heavy-quark spin, $\vec{s}_{Q}$, with 
the ``light spin,'' $\vec{j}_{q} = \vec{L}+\vec{s}_{q}$. The good 
quantum numbers are then $L$, $j_{q}$, and $J$, where $\vec{J} = 
\vec{s}_{Q}+\vec{j}_{q}$, and the low-lying levels are
\begin{eqnarray*}
    L= 0: & j_{q} = \cfrac{1}{2}: & \begin{array}{c} 0^{-}\hbox{ - }1^{-} 
    \end{array} \\[2pt] 
    L = 1: & j_{q} = \left\{\begin{array}{c} \cfrac{1}{2}: \\[2pt] \cfrac{3}{2}: 
    \end{array}\right. & \begin{array}{c}
  0^{+}\hbox{ - }1^{+} \\[2pt]  1^{+}\hbox{ - }2^{+} \end{array}\; 
  \quad,\hbox{ etc.}   
\end{eqnarray*}
In the absence of configuration mixing, this classification implies 
that the $j_{q} = \cfrac{3}{2}$ states will decay only through the 
$d$-wave, and so will be narrow. The $j_{q} = \cfrac{1}{2}$ states, 
for which $s$-wave decay is allowed, will in general be broad. 
It bears emphasis that the $D_{s}$, $B_{s}$ systems could be exceptions 
to this rule, because of the limited phase space available for kaon 
emission.

It makes sense to seek out intermediate cases wherever we can find 
them. We expect, for example, mixed $1^{+}$ levels in the $B_{c} = 
b\bar{c}$ spectrum, but detailed information is not likely to be in our hands 
soon. A more accessible case might be that of the strange particles 
($s\bar{q}$), for which the $q\bar{q}$-inspired $LS$  
classification has been the standard. Perhaps some unexpected 
insights might come from considering strange mesons as heavy-light
($Q\bar{q}$) states~\cite{Eichten:1993ub}. In any event, it is worth 
asking how infallible is the intuition we derive from regarding $D_{s}$ 
states as heavy-light.

Here in Lisbon, we heard an indication from the Belle experiment that
configuration mixing may not be negligible for the $1^{+}$ $D_{s}$
levels.  An angular analysis of the decay $D_{s1}(2536) \to
D^{*+}K_{S}$ indicates the presence of a large $s$-wave
amplitude~\cite{drutskoy}.  That is to say, the putative
$j_{\ell}=\cfrac{3}{2}$ state seems not to decay in a pure $d$-wave. 
Nevertheless, $D_{s1}(2536)$ remains narrow, with a total width less 
than $2.3\mev$. Is this because the $j_{\ell}=\cfrac{1}{2}$ level 
with which it might mix is anomalously narrow (as we shall recall 
next), or is there another explanation for the small $s$-wave width?

Two states that might be identified as the $j_{q} = \cfrac{1}{2}$
$c\bar{s}$ levels are well established, the $0^{+}\; D_{sJ}^{*}(2317)
\to D_{s}\pi^{0}$ and $1^{+} \;D_{sJ}(2460) \to D_{s}\gamma,
D_{s}^{*}\pi^{0}$.  Their centroid lies some $135\mev$ below that of
the $j_{\ell}=\cfrac{3}{2}$ states, the $1^{+}~D_{s1}(2536)$ and
$2^{+}~D_{s2}^{*}(2573)$.  The low masses disagree with relativistic
potential model predictions, and mean that the expected strong decay by
kaon emission is kinematically forbidden.

The fact that $D_{sJ}^{*}(2317)$ and $D_{sJ}(2460)$ appeared in
isospin-violating decays stimulated interpretations beyond the standard
$c\bar{s}$ body plan, including $DK$ molecules and tetraquarks.  It is
noteworthy that the BaBar experiment looked for, but did not find,
partners with charge $0$ or $\pm2$~\cite{Aubert:2004bp}.  Radiative decay
rates should be an incisive diagnostic~\cite{colangelo}.  For any
interpretation of the $D_{sJ}$ states, it is imperative to predict what
happens in the $B_{s}$ system.  Experimenters need not wait for the
theorists to place their bets.  Tracking down the $B_{sJ}$ analogues
should be a high priority for CDF and D\O!

I think the evidence is persuasive that the $D_{sJ}$ levels are
ordinary $c\bar{s}$ states at lower masses than anticipated, and I find
it intriguing that these states might give us a window on chiral
symmetry in a novel
setting~\cite{Bardeen:2003kt,Nowak:2003ra,Kalashnikova:2005tr}.
Let us suppose that, contrary to standard intuition in light-quark
systems, chiral symmetry and confinement might coexist in heavy--light
mesons.  Then we would expect to observe chiral supermultiplets: states
with orbital angular momenta $L, L+1$, but the same value of $j_{q}$.
Specifically, we should find the paired doublets
    \begin{displaymath}
	\begin{array}{l}
	  j_{q} = \cfrac{1}{2}:  1\mathrm{S}(0^{-},1^{-})\hbox{ and } 
	1\mathrm{P}(0^{+},1^{+});  \\[6pt]
		      j_{q} = \cfrac{3}{2}:  
		   1\mathrm{P}(1^{+},2^{+})\hbox{ and } 
	1\mathrm{D}(1^{-},2^{-}). 
	\end{array}
    \end{displaymath}
Chiral symmetry predicts equal hyperfine splitting in the paired
doublets, $M_{D_{s}(1^{+})} - M_{D_{s}(0^{+})} = M_{D_{s}(1^{-})} -
M_{D_{s}(0^{-})}$, in agreement with what is observed. So far, 
the predictions for decay rates match 
experiment~\cite{Eichten:2005sf,estiabeijing}. In addition to 
confronting chiral symmetry's predictions for the $D_{s}$ and other 
families, we need to ask to what extent the coexistence of chiral 
symmetry and confinement is realized in QCD, and how chiral symmetry 
may be restored in excited states~\cite{Swanson:2003ec}.

\section{Quarkonium Spectroscopy}

In the parallel sessions, we had the pleasure of hearing a flood of
beautiful new results on $\psi$ and $\Upsilon$ spectroscopy. Here are 
some of the highlights.

The CLEO experiment reported the discovery of the long-sought $h_{c}
{(1\slj{1}{2}{1})}$level in $\psi^{\prime} \to
\pi^{0}h_{c}$~\cite{miller}.  The mass of the new state, $M(h_{c}) =
3524.4 \pm 0.6 \pm 0.4\mev$, is about $1\mev$ below the 1\slj{3}{2}{J}
centroid.  Belle reported extensive studies of $\gamma\gamma \to
\eta_{c}, \chi_{c0}, \chi_{c2} \to h^{+}h^{-},
h^{+}h^{-}h^{+}h^{-}$~\cite{sokolov}.  Several of the rates
$\Gamma(\eta_{c} \to \gamma\gamma)\mathcal{B}(\eta_{c} \to f)$ are
about one-third of current world averages~\cite{Eidelman:2004wy}.  CLEO
has observed the rare decay $\psi(3770) \to \pi\pi\,\jpsi$ at a
branching fraction $\mathcal{B}(\psi(3770) \to \pi^{+}\pi^{-}\,\jpsi) =
(214 \pm 25 \pm 22)\times 10^{-5}$~\cite{Adam:2005mr}.  This is important
engineering information for anticipating the properties of the
1\slj{1,3}{3}{2} levels.  They have identified a rare radiative decay
of $\psi^{\prime\prime}$, with a partial width $\Gamma(\psi(3770) \to
\gamma \chi_{c1}) = 75 \pm 18\kev$~\cite{Coan:2005ps}.  Finally---in the charmonium
sector---the KEDR experiment in Novosibirsk has employed resonance
depolarization techniques to make precise energy determinations that
enable them to characterize the masses $M(\psi^{\prime}) = 3686.117 \pm
0.012 \pm 0.015\mev$ and $M(\psi^{\prime\prime})= 3773.5 \pm 0.9 \pm
0.6\mev$~\cite{todyshev}.

In the $b\bar{b}$ sector, CLEO has presented another measurement of
great engineering significance, the determination of the $B_{s}^{(*)}$
yield on the 5S resonance: $\mathcal{B}(\Upsilon(5S) \to
B_{s}^{(*)}\bar{B}_{s}^{(*)}) = 16.0 \pm 2.6 \pm 6.3 \%$~\cite{Artuso:2005xw}, and has made
a precise determination of the 1S, 2S, and 3S leptonic
widths~\cite{duboscq}.  One number that shows the quality of the
measurements is $\Gamma(\Upsilon(1\hbox{S}) \to e^{+}e^{-}) = 1.336 \pm
0.009 \pm 0.019\kev$.  Such information will provide a good test of
lattice calculations of the bottomonium spectrum~\cite{Gray:2005ur} and provides needed
input for improved potential-model descriptions.  There is also news
about hadronic cascades.  The Belle experiment has observed $38 \pm
6.9$ events consistent with the transition $\Upsilon(4\hbox{S}) \to
\pi^{+}\pi^{-}\Upsilon(1\hbox{S})$, corresponding to a branching
fraction $\mathcal{B}= (1.1 \pm 0.2 \pm 0.4)\times
10^{-4}$~\cite{sokolov}.  CLEO has offered evidence for the first
observation of  hadronic cascades not involving a \slj{3}{1}{1} level,
determining  partial widths $\Gamma \approx 0.9\kev$ for the transitions
$\chi_{b}^{\prime}(2\slj{3}{2}{2,1}) \to
\pi^{+}\pi^{-}\chi_{b}(1\slj{3}{2}{2,1})$~\cite{duboscq}.

\section{New States Associated with Charmonium}
In the three years since the Belle Collaboration announced the
observation of the 2\slj{1}{1}{0} $c\bar{c}$ state in exclusive $B$
decays~\cite{Choi:2002na}, new states have arrived in great profusion.
In addition to the 1\slj{1}{2}{1} $h_{c}(3524)$ level already
mentioned~\cite{Rosner:2005ry}, we have credible evidence for five new
particles connected with the charm-anticharm system.  The best known of
these is $X(3872)$~\cite{pakhlov}, clearly established in several
experiments.  On current evidence, it is likely to be a $J^{PC} =
1^{++}$ state, and is \textit{probably not charmonium.} More about the
$X(3872)$ shortly.

The remaining particles need confirmation; each has been seen only in a
single experiment so far.  Belle~\cite{pakhlov} has reported $Y(3940
\pm 11)$ in the decay $B \to K\omega\jpsi$; it is a relatively broad
state, with a total width $\Gamma = 92 \pm 24\mev$.  Belle also reports
the state $X(3936 \pm 14)$, seen in $e^{+}e^{-} \to \jpsi +
X$~\cite{Abe:2005hd}.  It is observed to decay into $D\bar{D}^{*}$, but
not $D\bar{D}$, which suggests an unnatural parity assignment.  The
total width is $\Gamma = 39 \pm 24\mev$.  These characteristics make
$X(3936)$ a plausible $\eta_{c}^{\prime\prime} (3\slj{1}{1}{0})$
candidate~\cite{Rosner:2005gf,Eichten:2005ga}.

Belle has observed a narrow ($\Gamma \approx 20\mev$) state in $\gamma\gamma 
\to D\bar{D}$  that they call $Z(3931 \pm 4 \pm 2)$~\cite{Abe:2005bp}. 
The production and decay characteristics are consistent with a 
$2^{++}$ assignment, and this state is a plausible $\chi_{c2}^{\prime}$ 
(2\slj{3}{2}{2}) candidate~\cite{Rosner:2005gf,Eichten:2005ga}. The most recent 
addition to the collection is $Y(4260)$, a $1^{--}$ level seen by 
BaBar in $e^{+}e^{-} \to \gamma 
\pi^{+}\pi^{-}\,\jpsi$~~\cite{Aubert:2005rm}, with supporting 
evidence from $B \to K^{-}\jpsi\pi\pi$~\cite{Aubert:2005zh}.

As if seven new states were not enough, there are more charmonium
levels to be found~\cite{Eichten:2004uh,Barnes:2005pb}.  Two of
these---the unnatural parity 1\slj{1,3}{3}{2} states that should lie
between $D\bar{D}$ and $D\bar{D}^{*}$ threshold---have been anticipated
for three decades.  The $2^{--}$ $\psi_{2}(3831)$ (1\slj{3}{3}{2})
state should be seen to decay into $\gamma \chi_{c1,2}$ and
$\pi\pi\jpsi$, but not to $D\bar{D}$.  Its hyperfine partner, the
$2^{-+}$ $\eta_{c2}(3838)$ {(1\slj{1}{3}{2})}, should be observed in
decays to $\gamma h_{c}$ and $\pi\pi\eta_{c}$, but not to $D\bar{D}$.
Then there are a couple of states, along with the 2\slj{3}{2}{2} and
perhaps 3\slj{1}{1}{0} levels mentioned above, that we have only come
to anticipate as narrow on the basis of recent coupled-channel
calculations.  These are the $3^{--}$ $\psi_{3}(3868)$
{(1\slj{3}{3}{3}), which should be observed as a quite narrow ($\Gamma
\ltap 1\mev$) peak in $D\bar{D}$, and the $4^{++}$ $\psi_{4}(4054)$
(1\slj{3}{4}{4}), which should also be seen as a $D\bar{D}$ resonance
with $\Gamma \ltap 5 \mev$.  And let us not forget the possibility that
gluonic degrees of freedom will manifest themselves in the form of
hybrid $c\bar{c}g$ levels~\cite{Liao:2002rj,petrov,Close:2005iz}.

All of these $X$s and $Y$s are very confusing, so we may have to admit
that our alphabet is not rich enough to accommodate the new reality of
charmonium spectroscopy.  By good fortune, Dr.~Seuss, author of the
children's classic, \textit{O Gato do Chap\'{e}u,} has anticipated our
need and extended the latin alphabet to include new
letters such as \textsf{quan, yekk, spazz,} and \textsf{floob}~\cite{seuss}. 
Should we assign to the Particle Data Group
the responsibility of deciding which particle is a \textsf{yekk} and which a 
\textsf{floob,} or should that honor rest with the discoverers?

\section{What is $X(3872) \to \pi\pi\,\jpsi$?}
The $X(3872)$ is the best studied of the new $c\bar{c}$-associated
states, and it has been subjected to a broad range of diagnostic tests.
Upon discovery, $X(3872)$ seemed a likely candidate for $\psi_{2}$ (or
perhaps $\psi_{3}$), but the expected radiative transitions to
$\chi_{c}$ states have never been seen.  The $\pi\pi$ mass spectrum
favors high dipion masses, suggesting a $\jpsi\,\rho$ decay that is
incompatible with the identification of $X(3872) \to
\pi^{+}\pi^{-}\,\jpsi$ as the strong decay of a pure isoscalar state.
Observing---or limiting---the $\pi^{0}\pi^{0}\,\jpsi$ decay remains an
important goal.  An observed $\jpsi\,3\pi$ decay suggests an
appreciable transition rate to $\jpsi\,\omega$.  Belle's $4.4$-$\sigma$
observation of the decay $X(3872) \to \jpsi\,\gamma$ determines $C=+$,
opposite to the charge-conjugation of the leading charmonium
candidates.  Finally, an analysis of angular distributions supports the
assignment $J^{PC} = 1^{++}$, but the mass of $X(3872)$ is too low to
be gracefully identified with the 2\slj{1}{2}{1} charmonium state,
especially if $Z(3931)$ is identified as the 2\slj{3}{2}{2} level. 
[It is important to note that our expectations for charmonium states 
above $D\bar{D}$ threshold have matured to include the coupling of 
$c\bar{c}$ levels with open charm.]

If $X(3872)$ is not a charmonium level, what might it be?  Three
interpretations take the near-coincidence of the new state's mass and
the $D^{0}\bar{D}^{*0}$ to be a decisive clue: an $s$-wave cusp at
$D^{0}\bar{D}^{*0}$ threshold~\cite{Bugg:2004rk}, a $D^{0}$ --
$\bar{D}^{*0}$ ``molecule'' bound by pion
exchange~\cite{Tornqvist:2004qy,Close:2003sg,swanghp,Voloshin:2003nt}, and a
diquark--antidiquark ``tetraquark'' state
$[cq][\bar{c}\bar{q}]$~\cite{piccinini,Maiani:2004vq,Maiani:2005pe}. 
What distinctive predictions might allow us to put these 
interpretations to the test? On the threshold enhancement 
interpretation, we should expect bumps at many thresholds, but no 
radial or orbital excitations. If pion exchange is decisive, then 
there should be no analogue molecule at $D_{s}\bar{D}_{s}^{*}$ 
threshold. The tetraquark interpretation suggests that $X(3872)$ 
should be split into two levels, because $[cu][\bar{c}\bar{u}]$ and 
$[cd][\bar{c}\bar{d}]$ would be displaced by about $7\mev$. If 
diquarks are useful dynamical objects, there should be a sequence of 
excited states as well.   

The implication that $X(3872)$ could be resolved into two states has
already attracted experimental attention from
BaBar~\cite{Aubert:2005zh}.  The evidence is far from decisive, but I
report it to you as an illustration of the lively dialogue between
experiment and theory that has characterized this subject. 
$61.2 \pm 15.3$ events that fit the hypothesis $B^{-} \to 
K^{-}X(3872)$ lead to a mass of $3871.3 \pm 0.6 \pm 0.1\mev$, whereas 
$8.3 \pm 4.5$ $B^{0} \to K^{0}X(3872)$ events yield $3868.6 \pm 1.2 \pm 
0.2\mev$. The mass difference, $2.7 \pm 1.3 \pm 0.2\mev$, doesn't yet 
distinguish between one $X$ and two. The same study compares the ratio 
of the charged and neutral decays,
$\mathcal{R} \equiv {\mathcal{B}(B^{0} \to
K^{0}X(3872))}/{\mathcal{B}(B^{-} \to K^{-}X(3872))} = 0.50 \pm 0.30 \pm
0.05$,
to be compared with the expectations of the tetraquark ($\mathcal{R} \approx 
1$) and molecule ($\mathcal{R} \ltap 0.1$) pictures.

Braaten \& Kusunoki~\cite{Braaten:2003he} have called attention to a
fascinating phenomenon (known in nuclear physics as a Feshbach
resonance) that should occur if a dynamical level and a threshold
coincide: an extremely large scattering length that is governed
(inversely) by the difference between the bound-state energy and the
threshold.  I do not think that $X(3872)$ meets the conditions, but we
should be attentive for this circumstance---perhaps even for one of the
other new states.

The campaign to understand $X(3872)$ has called on numerous heuristic
pictures, and has spurred theorists to elaborate simple images into
calculational tools.  Coupled-channel potential models appear to be
useful interpretive tools; they have help us learn what $X(3872)$ is
not, and we will see how helpful they can be in making sense of the
other new states.  One can only be impressed with the increasing
effectiveness of lattice QCD (below
threshold)~\cite{davies,diPierro:2003bu,Gray:2005ur}.  We still await a
definitive sighting of the influence of the gluonic degrees of freedom
on the spectrum of quarkonium or states related to quarkonium.  To test
our understanding of $X(3872)$ and the other new states, it would be
extremely helpful to know what happens in the $b\bar{b}$ system. For 
a more extensive recent discussion of the new states in the 
charmonium system, see Ref.~\cite{Swanson:2005tq}.

\section{Outlook{\protect \footnote{See Ref.~\cite{Close:2004ip} for a 
different emphasis and more expansive view of the subject.}}}
Hadronic physics is rich in opportunities.
Models---disciplined by principles---are wonderful exploratory tools
that can help us to uncover regularities and surprises.  It is
important that phenomenological studies make contact at every
opportunity with symmetries and with lattice QCD, especially as the
incorporation of dynamical quarks becomes routine.  Our goal---it is
the goal of all science---must be to build coherent networks of
understanding, not one-off interpretations of data.  In both experiment
and theory, in both exploration and explanation, we profit by tuning
between systems with similar but not identical characteristics, and by
driving models beyond their comfort zones.

In spectroscopy, I see much to be gained from a comparison of the
hadronic body plans we know: quark--antiquark mesons and three-quark
baryons, with the diversity that springs from light and heavy quarks.
Light-quark mesons, heavy-light mesons, and heavy quarkonia call upon
different elements of our theoretical armamentarium, as do baryons
containing 3, 2, 1, or 0 light quarks---but all are hadrons, and some
of what we learn in one setting should serve us in another.  Do other
body plans occur in Nature---two-quark--two-antiquark mesons,
four-quark--one-antiquark baryons, and more?  What r\^{o}le do diquarks
play in determining the hadron spectrum and interactions?  And what
lessons might we draw from the behavior of hadronic matter under
unusual conditions, including those that prevail in heavy-ion
collisions?

High-rate experiments more incisive than ever before are 
giving us new looks at familiar phenomena and new opportunities to 
exploit established techniques. Dalitz-plot analyses offer exquisite 
sensitivity to small amplitudes and access to phase information. We 
are gaining a richer understanding of diffraction, hadronization, and 
the structure of the proton.

In addition to the specific measurements I have mentioned and that 
others have highlighted in the course of this meeting, I would like to 
underscore the value of broad searches for new mesons and baryons. 
BaBar's discovery of $D_{sJ}$ and Belle's string of observations remind 
us that you don't have to know precisely what you are looking for to 
find something interesting: combining a convenient trigger particle with an
identifiable hadron or two---$(\jpsi\hbox{ or }\Upsilon) + \pi, \pi\pi,
K, K_{S}, p, \Lambda, \gamma, \eta, \omega, \ldots$---can be very
profitable indeed.

In experiment and theory alike, let us use our models and our
truncated versions of QCD to guide our explorations and organize our
understanding.  Let us keep in mind the limitations of our tools as we
focus on what we can learn of lasting value.  Let us, above all, try to
discern where the real secrets are hidden.	

\begin{acknowledgments}
 I am grateful to Pepe Bernab\'{e}u and the Organizing Committee for their
 kind invitation to present this report, and for the stimulating and
 enjoyable atmosphere of HEPP2005.  I thank Helen Quinn for her
 thought-provoking comments on hadronic physics, delivered at
 \textit{Heavy Quarks and Leptons 2004}. Andreas Kronfeld called my 
 attention to Ref.~\cite{seuss}. Jon Rosner offered helpful advice.
 
It is my pleasure to thank Karin Daum and Gerhard Mallot for their 
insightful work as convenors of the Hadronic Physics parallel sessions, 
and to acknowledge the important contributions of the forty-three 
speakers. I regret that I could not discuss every contribution in 
this short summary.
\end{acknowledgments}


\begin{thebibliography}{99}
    \bibitem{Veneziano:1968yb}
      G.~Veneziano,
      Nuovo Cim.\ A {\bf 57}, 190 (1968).

      \bibitem{Wilczek:1999be}
F.~ Wilczek, {Phys. Today}{ \bf 52 \textnormal{(11)}} 11 (November 1999).

\bibitem{alltalks} Parallel session talks are available at 
\urll{http://www.lip.pt/events/2005/hep2005/talks/session5.html}.

\bibitem{heller} U.~M.~Heller, ``Some results from full 2+1 flavor simulations 
of QCD,'' \had.

\bibitem{ukawa} A. Ukawa, ``Hadron Spectrum from Lattice,'' talk 
 at International Conference on QCD and Hadronic Physics, Beijing, 
 June 2005,
\urll{http://www.phy.pku.edu.cn/~qcd/transparency/20-plen-m/Ukawa.pdf}.  

\bibitem{Manohar:1983md}
A.~Manohar and H.~Georgi 
\newblock {Nucl. Phys.}{ \bf B234,} 189 (1984).

\bibitem{Eichten:1991mt}
E.~J.~Eichten, I.~Hinchliffe, and C.~Quigg,
\newblock {Phys. Rev. D}{ \bf 45,} 2269 (1992).

\bibitem{Ellis:1973kp}
  J.~R.~Ellis and R.~L.~Jaffe,
  Phys.\ Rev.\ D {\bf 9}, 1444 (1974)
  [Erratum-ibid.\ D {\bf 10}, 1669 (1974)].

\bibitem{pretz} J.~Pretz, ``Spin structure with COMPASS,'' \had.

\bibitem{procureur} S.~Procureur, ``Recent measurement of $\Delta G/G$ at 
COMPASS,'' \had.

\bibitem{avetisyan} E.~Avetisyan, ``Transverse spin effects in single and 
double hadron production at Hermes,'' \had.

\bibitem{seuster} R.~Seuster, ``Fragmentation functions at Belle,'' 
\had.

\bibitem{davies} C.~Davies, ``Non-perturbative Field Theory,'' plenary 
  talk at HEPP2005, 
  \href{http://www.lip.pt/events/2005/hep2005/talks/hep2005_talk_ChristineDavies.pdf}{http://www.lip.pt/events/2005/hep2005/talks/hep2005\_talk\_ChristineDavies.pdf}.

  \bibitem{Hughes:1983kf}
  V.~W.~Hughes and J.~Kuti, 
  \newblock {Ann. Rev. Nucl. Part. Sci.}{ \bf 33,} 611 (1983).

  \bibitem{Melnitchouk:1995fc}
  W.~Melnitchouk and A.~W.~Thomas,
  \newblock {Phys. Lett.}{ \bf B377,} 11 (1996)
  \newblock [arXiv:nucl-th/9602038].

  \bibitem{Anselmino:1992vg}
    M.~Anselmino,
    E.~Predazzi, S.~Ekelin, S.~Fredriksson and D.~B.~Lichtenberg,
    Rev.\ Mod.\ Phys.\  {\bf 65}, 1199 (1993).


  \bibitem{Jaffe:1976ig}
  R.~L.~Jaffe,
  \newblock {Phys. Rev. D}{ \bf15,} 267 (1977).

  \bibitem{Brodsky:2004hh}
    S.~J.~Brodsky,
    Acta Phys.\ Polon.\ B {\bf 36}, 635 (2005)
    [arXiv:hep-ph/0411028].
   
  \bibitem{Jaffe:2003sg}
  R.~L.~Jaffe and F.~Wilczek,
  \newblock {Phys. Rev. Lett.}{ \bf 91,} 232003 (2003)
  \newblock [arXiv:hep-ph/0307341].

\bibitem{Shuryak:2005pk}
  E.~V.~Shuryak,
  J.\ Phys.\ Conf.\ Ser.\  {\bf 9}, 213 (2005)
  [arXiv:hep-ph/0505011].

\bibitem{Jenkins:2004tm}
  E.~Jenkins and A.~V.~Manohar,
  Phys.\ Rev.\ Lett.\  {\bf 93}, 022001 (2004)
  [arXiv:hep-ph/0401190].

\bibitem{selem} A. Selem, \textit{A Diquark Interpretation of the Structure and
Energies of Hadrons,} S. B. Thesis, MIT, 2005; see
F. Wilczek, ``Diquarks as inspiration and as objects,'' in
\textit{From Fields to Strings: Circumnavigating Theoretical Physics,} 
Ian Kogan Memorial Collection
edited by M.~Shifman, A.~Vainshtein, and J.~Wheater (World Scientific, 
Singapore, 2005) vol.~1, pp.~77-93
[arXiv:hep-ph/0409168].

  \bibitem{Cristoforetti:2004kj}
    M.~Cristoforetti, P.~Faccioli, G.~Ripka and M.~Traini,
    Phys.\ Rev.\ D {\bf 71}, 114010 (2005)
    [arXiv:hep-ph/0410304].
    
  \bibitem{'tHooft:1973jz}
  G.~'t~Hooft,
  \newblock {Nucl. Phys.}{ \bf B72,} 461 (1974).

  \bibitem{Dashen:1993jt}
  R.~F.~Dashen, E.~Jenkins and A.~V.~Manohar,
    Phys.\ Rev.\ D {\bf 49}, 4713 (1994)
    [Erratum-ibid.\ D {\bf 51}, 2489 (1995)]
    [arXiv:hep-ph/9310379].

\bibitem{ebert} D.~Ebert, ``Relativistic description of heavy 
baryons,'' Heavy Flavours \had.

    \bibitem{Ebert:2005xj}
      D.~Ebert, R.~N.~Faustov and V.~O.~Galkin,
      Phys.\ Rev.\ D {\bf 72}, 034026 (2005)
      [arXiv:hep-ph/0504112].
 See also   D.~Ebert, R.~N.~Faustov, V.~O.~Galkin and A.~P.~Martynenko,
  Phys.\ Rev.\ D {\bf 66}, 014008 (2002)
  [arXiv:hep-ph/0201217], and
  D.~Ebert, V.~O.~Galkin and R.~N.~Faustov,
  Phys.\ Rev.\ D {\bf 57}, 5663 (1998)
  [Erratum-ibid.\ D {\bf 59}, 019902 (1999)]
  [arXiv:hep-ph/9712318].
  
  
    \bibitem{Jaffe:1976ih}
  R.~L.~Jaffe,
  Phys.\ Rev.\ D {\bf 15}, 281 (1977).

  \bibitem{Jaffe:2005md}
    R.~L.~Jaffe,
    Phys.\ Rev.\ D {\bf 72}, 074508 (2005)
    [arXiv:hep-ph/0507149].
    
    \bibitem{Karliner:2003dt}
      M.~Karliner and H.~J.~Lipkin,
      Phys.\ Lett.\ B {\bf 575}, 249 (2003)
      [arXiv:hep-ph/0402260].

\bibitem{Diakonov:1997sj}
D.~Diakonov,
in
\textit{Advanced School of non-perturbative quantum field physics : Proceedings,}
edited by  M.~Asorey and A.~Dobado (World Scientific, Singapore, 
1998), pp.~1-55
[arXiv:hep-ph/9802298].

\bibitem{Diakonov:1997mm}
 D.~Diakonov, V.~Petrov and M.~V.~Polyakov,
 Z.\ Phys.\ A {\bf 359}, 305 (1997)
 [arXiv:hep-ph/9703373].
	  
 \bibitem{Ellis:2004uz}
J.~R.~Ellis, M.~Karliner and M.~Praszalowicz,
JHEP {\bf 0405}, 002 (2004)
[arXiv:hep-ph/0401127].

\bibitem{burkert} V.~Burkert, ``Have pentaquark states been seen?'' 
plenary talk at Lepton-Photon 2005, Uppsala, 
\href{http://lp2005.tsl.uu.se/~lp2005/LP2005/programme/presentationer/8_burkert_LP_2005_talk_1.ppt}{http://lp2005.tsl.uu.se/~lp2005/LP2005/programme/presentationer/8\_burkert\_LP\_2005\_talk\_1.ppt}.

\bibitem{kabana} S.~Kabana, ``Review of pentaquark searches,'' High
Energy Nuclear Physics \had.

\bibitem{klausf} K.~Freudenreich, ``$\rho\rho$ production in two-photon 
collisions,'' \had. \\
I.~V.~Anikin, B.~Pire and O.~V.~Teryaev,
Phys.\ Lett.\ B {\bf 626}, 86 (2005)
[arXiv:hep-ph/0506277].



\bibitem{nakanobeijing} T.~Nakano, ``Search for Pentaquarks,''    
at the International Conference on QCD and Hadronic Physics, 
Beijing, June 2005,
\urll{http://www.phy.pku.edu.cn/~qcd/transparency/20-plen-m/Nakano.pdf}.

\bibitem{devita} R.~DeVita, ``Search for Pentaquarks at CLAS in 
Photoproduction from Proton,'' at April 2005 APS Meeting,
\href{http://www.jlab.org/div_dept/physics_division/talks/Background/Hall_B/DeVita_aps2005.ppt}{http://www.jlab.org/div\_dept/physics\_division/talks/Background/Hall\_B/DeVita\_aps2005.ppt}.

\bibitem{pukhaeva} N.~Pukhaeva, ``Search for Doubly Charged and Doubly Strange 
Pentaquarks states in $Z$-boson  decays with the DELPHI detector at 
LEP,'' \had.

\bibitem{zivko} T.~\v{Z}ivko, ``Search for Pentaquarks in Proton-Nucleus 
Collisions with HERA-B,'' \had.

\bibitem{chekanov} S.~Chekanov, ``Strange pentaquark production at 
HERA,'' \had.

\bibitem{ozerov} D.~Ozerov, ``Charmed pentaquark searches at 
HERA,'' \had.

\bibitem{grauges} E.~Graug\'{e}s. ``Pentaquark searches at BaBar,'' 
\had.

\bibitem{mizuk} R.~Mizuk, ``Search for the $\Theta^{+}$ pentaquark using kaon 
secondary interactions in the  material of the Belle detector,'' \had;
    K.~Abe \etal\ [Belle Collaboration],
  arXiv:hep-ex/0507014.

\bibitem{lund69} B.~Magli\v{c}, ``Meson Resonances,'' in 
\textit{Proceedings of the Lund International Conference on 
Elementary Particles,} held at Lund, Sweden, June 25--July 1, 1969, 
edited by G. F. von Dardel (Institute of Physics, Lund, 1969) 
pp.~269--320. See especially chapter 7.
For a brief post-mortem, see  P.~F.~Harrison,
  ``Blind analysis,''
  J.\ Phys.\ G {\bf 28}, 2679 (2002);
also in \textit{Advanced Statistical Techniques in Particle Physics,}
Edited by M.~R.~Whalley and L.~Lyons (Institute for Particle Physics 
Phenomenology, Durham, 2002)  
\urll{http://www.ippp.dur.ac.uk/Workshops/02/statistics/proceedings//harrison.pdf}.

\bibitem{selen} M.~Selen, ``Hadronic substructure \& Dalitz analyses at
CLEO,'' \had; for a Dalitz-plot animation, see
\href{http://www.lip.pt/events/2005/hep2005/talks/dalitz_animation.avi}{http://www.lip.pt/events/2005/hep2005/talks/dalitz\_animation.avi}.

\bibitem{altenburg} D.~Altenburg, ``Charm Spectroscopy at BaBar,'' 
\had.

\bibitem{gauzzi} P.~Gauzzi, ``Radiative Phi decays at KLOE,'' \had.

\bibitem{Eichten:1994gt}
  E.~J.~Eichten and C.~Quigg,
  Phys.\ Rev.\ D {\bf 49}, 5845 (1994)
  [arXiv:hep-ph/9402210].


\bibitem{Allison:2004be}
  I.~F.~Allison, C.~T.~H.~Davies, A.~Gray, A.~S.~Kronfeld, P.~B.~Mackenzie and J.~N.~Simone
		  [HPQCD Collaboration],
  Phys.\ Rev.\ Lett.\  {\bf 94}, 172001 (2005)
  [arXiv:hep-lat/0411027].

  \bibitem{Acosta:2005us}
    D.~Acosta {\it et al.}  [CDF Collaboration],
    ``Evidence for the exclusive decay $B_{c}^{\pm}\to \jpsi \,\pi^{\pm}$ and measurement of
    the mass of the $B_{c}$ meson,''
    arXiv:hep-ex/0505076.

\bibitem{Eichten:1993ub}
  E.~J.~Eichten, C.~T.~Hill and C.~Quigg,
  Phys.\ Rev.\ Lett.\  {\bf 71}, 4116 (1993)
  [arXiv:hep-ph/9308337].

  \bibitem{drutskoy} A.~Drutskoy, ``Charm hadronic physics at Belle,'' 
\had.

\bibitem{Aubert:2004bp}
  B.~Aubert {\it et al.}  [BABAR Collaboration],
  arXiv:hep-ex/0408067.

  \bibitem{colangelo} P.~Colangelo, ``On the structure of $D_{sJ}(2317)$ and 
$D_{sJ}(2460)$,'' \had.

\bibitem{Bardeen:2003kt}
  W.~A.~Bardeen, E.~J.~Eichten and C.~T.~Hill,
  Phys.\ Rev.\ D {\bf 68}, 054024 (2003)
  [arXiv:hep-ph/0305049].

  \bibitem{Nowak:2003ra}
    M.~A.~Nowak, M.~Rho and I.~Zahed,
    Acta Phys.\ Polon.\ B {\bf 35}, 2377 (2004)
    [arXiv:hep-ph/0307102].

\bibitem{Kalashnikova:2005tr}
  Y.~S.~Kalashnikova, A.~V.~Nefediev and J.~E.~F.~Ribeiro,
  Phys.\ Rev.\ D {\bf 72}, 034020 (2005)
  [arXiv:hep-ph/0507330].

  \bibitem{Eichten:2005sf}
E.~Eichten,
Nucl.\ Phys.\ Proc.\ Suppl.\  {\bf 142}, 242 (2005).

\bibitem{estiabeijing}
E.~J.~Eichten,
\newblock {``New Developments in Heavy-Light Systems,'' QWG3, Beijing,
  \href{http://www.qwg.to.infn.it/WS-oct04/WS3talks/Oct14-2/Eichten_Ds.pdf}{http://www.qwg.to.infn.it/WS-oct04/WS3talks/Oct14-2/Eichten\_Ds.pdf}}.

\bibitem{Swanson:2003ec}
L.~Y.~Glozman,
Acta Phys.\ Polon.\ B {\bf 35}, 2985 (2004)
[arXiv:hep-ph/0410194];
E.~S.~Swanson,
Phys.\ Lett.\ B {\bf 582}, 167 (2004)
[arXiv:hep-ph/0309296];
M.~Shifman,
``Highly excited hadrons in QCD and beyond,''
arXiv:hep-ph/0507246.


\bibitem{miller} D.~Miller, ``Charmonium at CLEO,'' \had.

\bibitem{sokolov} A.~Sokolov, ``Studies of conventional quarkonia at 
Belle,'' \had.

\bibitem{Eidelman:2004wy}
  S.~Eidelman {\it et al.}  [Particle Data Group],
  Phys.\ Lett.\ B {\bf 592}, 1 (2004).

  \bibitem{Adam:2005mr}
    N.~E.~Adam  [CLEO Collaboration],
    ``Observation of $\psi(3770) \to \pi\pi \jpsi$ and measurement of
    $\Gamma_{ee}(\psi(\mathrm{2S}))$,''
    arXiv:hep-ex/0508023.

    \bibitem{Coan:2005ps}
      T.~E.~Coan  [CLEO Collaboration],
      ``First observation of $\psi(3770) \to \gamma \chi_{c1} \to \gamma 
      \gamma \jpsi$,''
      arXiv:hep-ex/0509030.

  \bibitem{todyshev} K.~Todyshev, ``Precision measurements of masses of 
charmonium states,'' \had.

\bibitem{Artuso:2005xw}
  M.~Artuso  \textit{et al.} [CLEO Collaboration],
  ``First Evidence and Measurement of $B_s^{(*)} \bar{B}_s^{(*)}$ Production at the
  $\Upsilon(\mathrm{5S})$,''
  arXiv:hep-ex/0508047.

\bibitem{duboscq} J. Duboscq, ``$\Upsilon$ \& $\chi_{b}$ analyses at
CLEO,'' \had.

    \bibitem{Gray:2005ur}
A.~Gray, I.~Allison, C.~T.~H.~Davies, E.~Gulez, G.~P.~Lepage, J.~Shigemitsu and M.~Wingate,
``The $\Upsilon$ spectrum and $m_{b}$ from full lattice QCD,''
arXiv:hep-lat/0507013.


  \bibitem{Choi:2002na}
    S.~K.~Choi {\it et al.}  [BELLE collaboration],
    Phys.\ Rev.\ Lett.\  {\bf 89}, 102001 (2002)
    [Erratum-ibid.\  {\bf 89}, 129901 (2002)]
    [arXiv:hep-ex/0206002].

    \bibitem{Rosner:2005ry}
      J.~L.~Rosner {\it et al.}  [CLEO Collaboration],
      Phys.\ Rev.\ Lett.\  {\bf 95}, 102003 (2005)
      [arXiv:hep-ex/0505073];
      P.~Rubin \etal\  [CLEO Collaboration],
      arXiv:hep-ex/0508037.

\bibitem{pakhlov} P.~Pakhlov, ``New particles at Belle,'' \had.

\bibitem{Abe:2005hd}
  K.~Abe, \textit{et al.}
  ``Observation of a new charmonium state in double charmonium production 
  in $e^{+}e^{-}$ annihilation at $\sqrt{s} \approx 10.6\gev$,''
  arXiv:hep-ex/0507019.

\bibitem{Rosner:2005gf}
  J.~L.~Rosner,
  ``Hadron spectroscopy - a 2005 snapshot,''
  arXiv:hep-ph/0508155.


  \bibitem{Eichten:2005ga}
    E.~J.~Eichten, K.~Lane and C.~Quigg,
    ``New states above charm threshold,''
    arXiv:hep-ph/0511179.
  
  \bibitem{Abe:2005bp}
  K.~Abe, \textit{et al.}
    ``Observation of a $\chi_{c2}^{\prime}$ candidate in $\gamma \gamma 
    \to D \bar{D}$ production in Belle,''
    arXiv:hep-ex/0507033.
    
    \bibitem{Aubert:2005rm}
      B.~Aubert {\it et al.}  [BABAR Collaboration],
      ``Observation of a broad structure in the $\pi^{+} \pi^{-}\jpsi$ 
      mass spectrum around
      $4.26\gevcc$,''
      arXiv:hep-ex/0506081.

      \bibitem{Aubert:2005zh}
B.~Aubert  \etal\ [BABAR Collaboration],
``Study of $\jpsi \pi^{+} \pi^{-}$ states produced in $B^{0} 
\to \jpsi \pi^{+} \pi^{-} K^{0}$ and $B^{-}
\to \jpsi \pi^{+} \pi^{-} K^{-}$,''
arXiv:hep-ex/0507090.

  \bibitem{Eichten:2004uh}
  E.~J.~Eichten, K.~Lane and C.~Quigg,
  Phys.\ Rev.\ D {\bf 69}, 094019 (2004)
  [arXiv:hep-ph/0401210].

  \bibitem{Barnes:2005pb}
    T.~Barnes, S.~Godfrey and E.~S.~Swanson,
    Phys.\ Rev.\ D {\bf 72}, 054026 (2005)
    [arXiv:hep-ph/0505002].

  \bibitem{Liao:2002rj}
  K.~J.~Juge, J.~Kuti and C.~J.~Morningstar,
  Phys.\ Rev.\ Lett.\  {\bf 90}, 161601 (2003)
  [arXiv:hep-lat/0207004].
    X.~Liao and T.~Manke,
    ``Excited charmonium spectrum from anisotropic lattices,''
    arXiv:hep-lat/0210030.

  \bibitem{petrov} A.~A.~Petrov, 
  \href{http://www.iop.org/EJ/article/1742-6596/9/1/013/jpconf5_9_013.pdf}
  {J. Phys.: Conf. Ser. \textbf{9,} 83-86 (2005)}.
	
  \bibitem{Close:2005iz}
  S.~L.~Zhu,
  Phys.\ Lett.\ B {\bf 625}, 212 (2005)
  [arXiv:hep-ph/0507025].
  F.~E.~Close and P.~R.~Page,
  Phys.\ Lett.\ B {\bf 628}, 215 (2005)
  [arXiv:hep-ph/0507199].

\bibitem{seuss} T.~S.~Geisel (Dr.~Seuss),  \textit{On Beyond 
Zebra} (Random House, New York, 1955). For images of the new 
characters, consult 
\urll{http://www.evertype.com/standards/csur/seuss.html}. 

\bibitem{Bugg:2004rk}
  D.~V.~Bugg,
  Phys.\ Lett.\ B {\bf 598}, 8 (2004)
  [arXiv:hep-ph/0406293]; 
  Phys.\ Rev.\ D {\bf 71}, 016006 (2005)
  [arXiv:hep-ph/0410168].
  See also 
  Phys.\ Rept.\  {\bf 397}, 257 (2004)
  [arXiv:hep-ex/0412045].

  \bibitem{Tornqvist:2004qy}
  N.~A.~{T\o rnqvist}
  \newblock {Phys. Lett.}{ \bf B590,} 209 (2004)
  \newblock [arXiv:hep-ph/0402237].

  \bibitem{Close:2003sg}
    F.~E.~Close and P.~R.~Page,
    Phys.\ Lett.\ B {\bf 578}, 119 (2004)
    [arXiv:hep-ph/0309253].
    
\bibitem{swanghp} E.~S.~Swanson, 
\href{http://www.iop.org/EJ/article/1742-6596/9/1/012/jpconf5_9_012.pdf}
{J. Phys.: Conf. Ser. \textbf{9,} 79-82 (2005)}.

\bibitem{Voloshin:2003nt}
  M.~B.~Voloshin,
  Phys.\ Lett.\ B {\bf 579}, 316 (2004)
  [arXiv:hep-ph/0309307].

\bibitem{piccinini} F.~Piccinini, ``Diquark-antiquarks states with hidden 
or open charm,'' \had.
 
\bibitem{Maiani:2004vq}
  L.~Maiani, F.~Piccinini, A.~D.~Polosa and V.~Riquer,
  Phys.\ Rev.\ D {\bf 71}, 014028 (2005)
  [arXiv:hep-ph/0412098].

  \bibitem{Maiani:2005pe}
    L.~Maiani, V.~Riquer, F.~Piccinini and A.~D.~Polosa,
    Phys.\ Rev.\ D {\bf 72}, 031502 (2005)
    [arXiv:hep-ph/0507062].

  \bibitem{Braaten:2003he}
    E.~Braaten and M.~Kusunoki,
    Phys.\ Rev.\ D {\bf 69}, 074005 (2004)
    [arXiv:hep-ph/0311147].

    \bibitem{diPierro:2003bu}
      M.~di Pierro {\it et al.},
      Nucl.\ Phys.\ Proc.\ Suppl.\  {\bf 129}, 340 (2004)
      [arXiv:hep-lat/0310042].
      
\bibitem{Swanson:2005tq}
E.~Swanson,
``Review of Heavy Hadron Spectroscopy,''
arXiv:hep-ph/0509327.

  \bibitem{Close:2004ip}
  F.~E.~Close,
  ``Hadron spectroscopy (theory): Diquarks, tetraquarks, pentaquarks and no
  quarks,''
    in \textit{ICHEP 2004,} Proceedings of the 32nd International 
    Conference on High-Energy Physics, edited by Hesheng Chen \etal\
    (World Scientific, Singapore, 2005), 
    vol.~1, pp.~100-107 
  [arXiv:hep-ph/0411396].


\end{thebibliography}
\end{document}